\documentclass[aps,prx,twocolumn,10pt,preprint,onecolumn]{revtex4-1} 

\usepackage{amssymb,amsmath}
\usepackage{graphicx}

\usepackage{multirow}
\usepackage{bm}

\usepackage[utf8]{inputenc}
\usepackage[T1]{fontenc}
\usepackage{mathptmx}

\begin{document}
	\title{Minimum entropy production effect on a quantum scale}

	\author{Ferenc M\'arkus}
	\email[Corresponding author: ]{markus@phy.bme.hu}
	\affiliation{Department of Physics,
		Budapest University of Technology and Economics,
		H-1111 Budafoki \'ut 8., Budapest, Hungary}

	\author{Katalin Gamb\'ar}
	\email[ ]{gambar.katalin@uni-obuda.hu}
	\affiliation{Institute of Microelectronics and Technology, \\ K\'alm\'an Kand\'o Faculty of Electrical Engineering, \\ \'Obuda University, \\ Tavaszmez\H{o} u. 17, H-1084 Budapest, Hungary}
    
    \affiliation{Department of Natural Sciences, \\ National University of Public Service, \\  Ludovika t\'er 2, H-1083 Budapest, Hungary}

	\date{\today}

\begin{abstract}
	Discovery of quantized electric conductance by the group of van Wees in 1988 was a major breakthrough in physics. Later, the group of Schwab has proven the existence of quantized thermal conductance. Advancing one step further, we present that quantized entropy current can be interpreted and it ease the description of a transferred quantized energy package. This might yield a universal transport behavior of the microscopic world. During the transfer of a single energy quantum, $h \nu$ between two neighboring domains the minimum entropy increment is calculated. Furthermore, the possible existence of the minimum entropy production can be formulated.
\end{abstract}

\maketitle

\section{Introduction}

Many concurrent research areas are relying the quantized conductances, such as electric \cite{Wees_1988}, thermal \cite{schwab2000,schwab2001,schwab2006}, integer \cite{Klitzing1980}, fractional \cite{tsui1982,laughlin1983} quantum Hall, and spin quantum Hall effect. Many recent studies focus on half-integer quantum Hall effect \cite{dora2015} in topological insulators, and in the quantized light–matter interaction on the edge state of a quantum spin Hall insulator \cite{gulacsi2016}. Furthermore, the electron transport through individual molecules as a candidate for future nanoelectronic circuits, also exhibits similar interesting properties \cite{dora2009,geresdi2011,geresdi2014}. \\
Based on the example of the quantized electric and thermal conductances the quantized entropy current can be introduced. This step enables us to incorporate the thermodynamic concepts into quantum processes. We show that the entropy change during the transfer of an energy package of $h \nu$ can be expressed. The description allows us to formulate a mathematical equation that expresses the minimal entropy production principle for the microscopic world. These results may be useful to acquire the maximal physical limits of reversibility of coherent quantum systems, e.g. in quantum dots and quantum computers relying on these. \\

The discussed theory incorporates various fields of physics. To aid the reader we give a brief summary of the used phenomena. Afterwards, a coherent frame arises in which the main aims and the applicable methods are discussed.

\section{Historical considerations}

\subsection{The quantized electric conductance}

\begin{figure}[h]
\centering
\includegraphics[width=8 cm, height=6 cm]{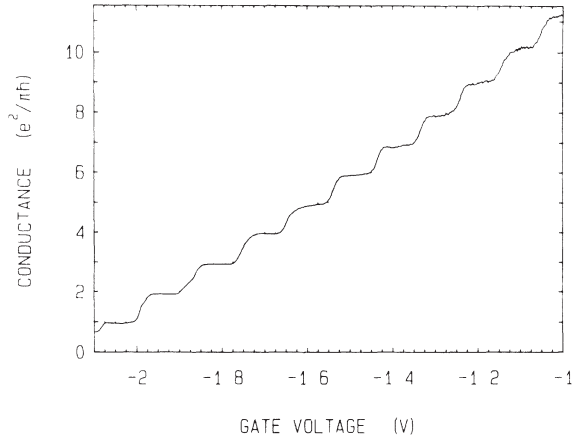}
\caption{The electric conductance as a function of gate voltage. The channel width, $w$, can be modulated by the gate voltage. The quantized behavior can be read out directly from the figure. The quantum of electric conductance is $2e^{2}/h$ based on theoretical predictions. Adopted from Reference \cite{Wees_1988} $\copyright$ (1988) American Physical Society.} \label{kvantalt_elektromos_vezetokepesseg}
\end{figure}

First, a short introduction to quantized electric conductance is given below. Landauer \cite{landauer_1957,landauer_1981,landauer_1989} theoretically predicted the existence and the amount of the quantized electric conductance which can be expressed as

\begin{equation}  \label{kvantalt_el_vez}
G = \frac{2e^2}{h} = 7.75 \cdot 10^{-5} \, \textrm{S},
\end{equation}

where $e$ is the elementary charge and $h$ is the Planck constant. In general 

\begin{equation}  \label{kvantalt_el_vez_kond}
G = R^{-1} = \sigma A/L,
\end{equation}

where $R$ is the ohmic resistance, $\sigma$ is the specific electric conductivity, $A$ is the cross section and $L$ is the length of the conductor. The experimental proof of the theoretical prediction was elaborated by van Wees \emph{et al} \cite{Wees_1988}. Theoretically, the quantized conductance may appear in a nanowire with width, $w$, comparable to the length of Fermi wavelength $\lambda_F$, and the length, $L$, is less than the free electron path. The quantized electric conductance was measured in a 2D electron gas realized within an AlGaAs–GaAs boundary layer. The gate voltage dependence of the electric conductance is shown in Fig. \ref{kvantalt_elektromos_vezetokepesseg}. 

Theoretically, the confined electrons, present in a long, straight 2D quantum wire can be described by the following wave function:

\begin{equation}
\Psi_{k,j}(x,y) \sim \sin{(ikx)} \sin{\left( \frac{j\pi}{w}y \right)},
\end{equation}

\noindent where $k$ is the wavenumber, $j$ is an integer quantum number. The first factor pertains to the propagating plane wave in the direction $x$, while the second factor describes the cross-modes in the direction $y$. The corresponding energy to the wave function $\Psi_{k,j}(x,y)$ expresses a continuum spectrum arising from the $x$-axis solution, while it contains quantized levels, due to the finite width $w$ in the $y$ direction \cite{nawrocki2008}:

\begin{equation}
\varepsilon (k,j) = \frac{{\hbar}^{2}k^{2}}{2m} + \frac{{\hbar}^{2}{\pi}^{2}}{2mw^{2}} j^{2}.
\end{equation}   

The quantized number of states $\varepsilon (k,j)$ below Fermi surface is $N \sim 2w/\lambda_F$. Moreover, if the thermal energy can be neglected to the chemical potential difference $\Delta\mu$ between the contacts, the electic current of the $j^{\textrm{th}}$ channel can be expressed as

\begin{equation}
I_j = e v_j \left( \frac{dn}{dE} \right)_j \Delta\mu = e^2 v_j \left( \frac{dn}{dE} \right)_j V,
\end{equation}

where $v_j$ is the propagation velocity along the direction y and ${dn}/{dE}_j$ is the density of states at the Fermi level for $j^{\textrm{th}}$ state, $V$ is the voltage difference $V = \Delta\mu / e$. The number of states between $k$ and $k + dk$ in one dimension for unit length is

\begin{equation}
\frac{dn}{dk} = \frac{1}{2\pi},
\end{equation}

by which the density of states can be obtained as

\begin{equation}
\left( \frac{dn}{dE} \right)_j = \left( \frac{dn}{dk} \frac{dk}{dE} \right)_j = \frac{2}{h v_j}, 
\end{equation}

where the factor $2$ arises from the spin degeneracy. The total current  

\begin{equation}
I = \sum_{j=1}^{N} I_j = \frac{2e^2}{h} N V
\end{equation}

can be expressed, where $N$ is the number of channels.  The quantized electric conductance can be read out from the formula.

\subsection{The quantized thermal conductance}

Based on termodynamic and information theory assumptions, Pendry --- analogously to the Landauer's formula --- intuitively predicted the expression of the maximal rate of cooling (transferable energy $Q$ per unit time giving the quantum limits for information flow) for one channel \cite{pendry1983} as

\begin{equation}  \label{pendry_flow}
\frac{dQ}{dt} \leq \frac{{\pi} k_{B}^{2} T^2}{3\hbar}, 
\end{equation}

\noindent where $k_B$ is the Boltzmann constant, $\hbar$ is the reduced Planck constant, and $T$ is the temperature. Dividing by $T$, the maximal entropy current ($dS/dt$) in one channel can be obtained as

\begin{equation}  \label{pendry_entropy}
\frac{dS}{dt} \leq \frac{{\pi} k_{B}^{2} T}{3\hbar}. 
\end{equation}

Later, Rego and Kirczenow \cite{rego1998} deduced the thermal conductance of a quantum wire using more sophisticated calculations. Their result is

\begin{equation}  \label{Rego_Kirczenow}
\Lambda = \frac{{\pi}^2 k_{B}^{2} T}{3h}. 
\end{equation}

\noindent Here, the $\Lambda$ notation is introduced for the quantum of thermal conductance. Comparing Eqs. (\ref{pendry_entropy}) and (\ref{Rego_Kirczenow}) a factor 2 difference can be noted. Furthermore, according to Eq. (6) it appears that the maximal entropy change and the quantized thermal conductance are related to each other. The origin of quantized thermal conductivity is explored by many theoretical groups from various viewpoints \cite{angelescu98,nishiguchi97,blencowe99,blencowe2004,li03}. \\ 

Other considerations based on the Drude-Lorentz theory also hints the existence of quantized thermal conductance. In the model, the relation between the $\lambda$ thermal conductivity and $\sigma$ electric conductivity can be expressed as

\begin{equation}
\lambda = \frac{{\pi}^{2}}{3} \left( \frac{k_{B}}{e} \right)^{2} T \sigma,
\end{equation}

\noindent where $\sigma$ is the specific electric conductivity \cite{solyom}. Substituting the half of the specific electric conductivity from Eq. (\ref{kvantalt_el_vez_kond}), following expression for the thermal conductivity can be derived as
 
\begin{equation}  \label{kvantalt_hovezeto_kepesseg} 
\lambda = \frac{{\pi}^{2} k_{B}^{2} T}{3h} \frac{L}{A}.
\end{equation}

\noindent This might look only a formal analogy, since the quantized thermal measurements were elaborated in semicondutors, while Drude-Lorentz model is formulated to describe conducting electrons. The factor 2 appears in the electric conductance $G=2e^{2}/h$ (Eq. (\ref{kvantalt_el_vez})) due to the two spin states of the electron. However, in semiconductors the phonons carry the heat conduction thus $G = e^{2}/h$ can be considered. Similarly to the electric conductance the thermal conductance 

\begin{equation}  \label{kvantalt_termikus_vezetokepesseg} 
\Lambda = \lambda \frac{A}{L} = \frac{{\pi}^{2} k_{B}^{2} T}{3h} = 9.46 \cdot 10^{-13} T \left[ \frac{\textrm{W}}{\textrm{K}} \right]
\end{equation}

can be obtained. This means that the thermal conductance has a close relation with the quantum of entropy current. \\ 

\begin{figure}[h]
\centering
\includegraphics[width=8 cm, height=6 cm]{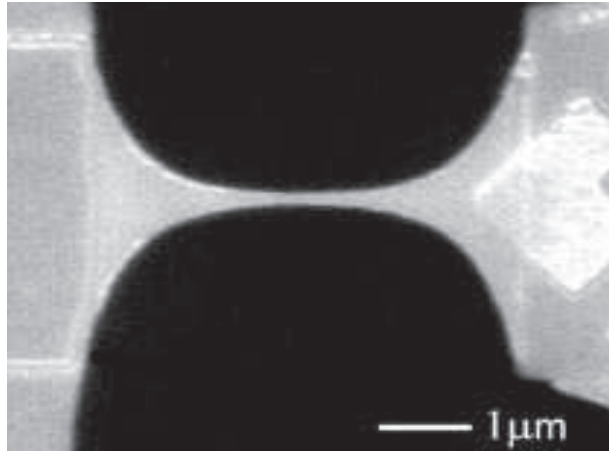}
\caption{Experimental realization of a Si$_{3}$N$_{4}$ waveguide \cite{schwab2000}. Physical dimensions: length: $L \sim 1 \, \mu$m; width: $w = 200$ nm; layer thickness: $d = 60$ nm. Adopted from Reference \cite{schwab2000} $\copyright$ (2000) Springer Nature.}  \label{termikus_hullamvezeto}
\end{figure}

The quantized thermal conductance was first measured in Si$_{3}$N$_{4}$ waveguide (see Fig. \ref{termikus_hullamvezeto}) by Schwab et al \cite{schwab2000}. Both the experimental setup and the measurement technique are fascinatingly sophisticated.

\begin{figure}[h]
\centering
\includegraphics[width=8 cm, height=6 cm]{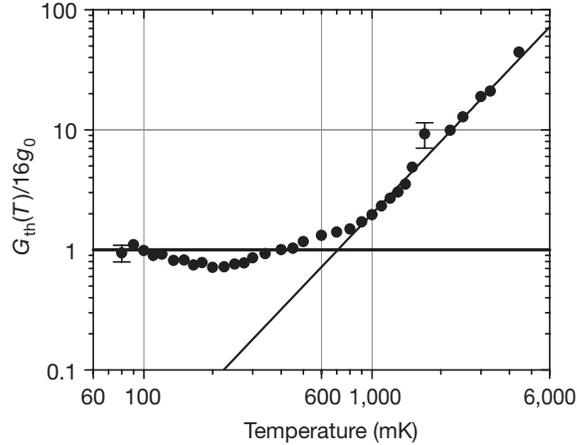}
\caption{The quantized behavior of thermal conductance \cite{schwab2000,schwab2001,schwab2006}. The appeared plato can be well recognized in the temperature range 60-700 mK. Adopted from Reference \cite{schwab2000} $\copyright$ (2000) Springer Nature.}  \label{termikus_vezetes_kvantum}
\end{figure}

The obtained result for the thermal conductivity in the temperature range of 60 mK to 6 K is shown in Fig. \ref{termikus_vezetes_kvantum}. The quantum of thermal conductivity based on theoretical considerations is
\begin{equation}
g_{0} = \frac{{\pi}^{2} k_{B}^{2} T}{3h}.
\end{equation}
Taking into account that in the measurement there four waveguides are present and each is expected to carry just four populated modes below the critical temperature of

\begin{equation}
T < T_{c} = \frac{\pi \hbar v}{k_B w} = 0.8 \, \text{K},
\end{equation}

the thermal conductance of the arrangement should approach a limiting value of $16 g_{0}$. In the measurement the width of the channel was $200$ nm, while $v = 6000$ m/s is the speed of sound in the material. The measurement data, normalized by $16 g_{0}$ \cite{schwab2000,schwab2001,schwab2006} is presented in Fig. \ref{termikus_vezetes_kvantum}. Please note, that below 700 mK data points significantly deviate from the linear fashion and converge to the value of $16 g_0$.

The constant thermal conductivity of $16g_0$ in the temperature range of $60-700$ mK proves the quantized behavior of the thermal conductance.

\subsection{Lagrangian description of heat conduction}

The quantum limit for information flow raised by Pendry is presumably related to an extreme-value problem. If so, it is probable that it is inherited from its thermodynamic background. To investigate, it is worth reviewing the extremum principle formulated for thermodynamics and the generating potential $\varphi$ introduced into the principle. The significance of these will be highlighted below.  \\

The theory is based on the least action principle

\begin{equation}
S_{\textrm{action}} = \int_{t_1}^{t_2} L dt = extremum,
\end{equation}

wherel $L$ is the Lagrange function of the problem. To proceed the principle has to be applied for the Fourier equation for heat conduction, which is a constant coefficient linear parabolic differential equation for temperature $T$ 

\begin{equation}  \label{foueq}
\varrho c_{v}\frac{{\partial}T}{{\partial}t} - {\lambda\nabla^2}T = 0,
\end{equation}

\noindent and which equation cannot be derived directly from the Hamiltonian principle for temperature $T$. Here, $\lambda$ is the thermal conductivity, introduced earlier, $c_v$ is the specific heat, and $\varrho$ denotes the mass density. Assume, that a potential space, $\varphi$, exists, which produces a measurable local equilibrium (classical) temperature field as follows:

\begin{equation}  \label{temp}
T(x,y,z,t) - T_0 = - \frac{{\partial}\varphi}{{\partial}t} - \frac{\lambda}{\varrho c_{v}} \nabla^2 \varphi = - \frac{{\partial}\varphi}{{\partial}t} - D \nabla^2 \varphi,
\end{equation}

\noindent where $T_0$ is a reference temperature. The presence of this reference temperature grants that the potential $\varphi$ has at least one well defined zero value and does not increase beyond all limits, e.g. limited from above. To simplify notation, it is worth introducing thermal diffusivity as

\begin{equation}
D = \frac{\lambda}{\varrho c_{v}}.
\end{equation}

\noindent Substituting the expression from Eq. \ref{temp} into the equation of heat conduction in Eq. (\ref{foueq}), the equation of motion of the problem can be obtained by the potential function of $\varphi$ as

\begin{equation}  \label{mozgasegy}
0 = - \frac{{\partial}^{2}\varphi}{{\partial}t^{2}}  + 
D^{2} \nabla^2 (\nabla^2 \varphi).
\end{equation}

\noindent The equation of motion (field equation) described in Eq. (\ref{mozgasegy}) is the Euler-Lagrange equation of heat conduction problem. The equation contains only self adjoint operators and thus it can be dedudced from the following Lagrangian

\begin{equation}  \label{lagrange}
L = \frac{1}{2} \left( \frac{{\partial}\varphi}{{\partial}t} \right)^{2} +
\frac{1}{2} D^{2} {({\nabla^2 \varphi})^{2}}
\end{equation}

\noindent \cite{mg,gm1994,gambar2016}. The presented method is also a good example of how the Hamilton's principle can be applied to dissipative processes \cite{szegleti2020}.\\

The Lagrangian theory can be quantized and the arising energy packages can be assigned to the thermal propagation \cite{markus95}. Performing the calculations for a silicon film with width 100 nm and cross section $10^{-6}$ m$^{2}$ at temperature $T=80$ mK the obtained first energy level is ${\epsilon}_{1}=7.0 \cdot 10^{-14}$ J = $4.375 \cdot 10^{5}$ eV \cite{vmg2009pre}. This energy value agrees well with the value of the transferred energy per unit time per unit temperature ${\epsilon}=7.6 \cdot 10^{-14}$ J = $4.75 \cdot 10^{5}$ eV calculated from the previously mentioned experimental results for for silicon nitride in Schwab et al \cite{schwab2000,schwab2001,schwab2006}. These results prove the material independence of quantized thermal conductance.

\section{The quantized behavior of the conductance of entropy current and the entropy production}

The change of an extensive physical quanity in a volume depends on the in- or outgoing flow via the total surface of the considered volume and the production or loss of the quantity within the volume. If the extensive quantity is the entropy, $S$, then the balance equation read as

\begin{equation}
\frac{dS}{dt} = -I_S + \Sigma ,
\end{equation}

where $I_S$ is the entropy current and $\Sigma$ is the entropy production. \\

Upon thermal propagation the relation between the entropy current density $J_s$ ($J_s = I_s / A$) and the heat current density ($J_q = - \lambda \nabla T$) can be expressed as usual. Applying Fourier's law yields 

\begin{equation}  \label{Fourier_tv}
J_q = -\lambda \nabla T .
\end{equation}

With the use of Eq. (\ref{Fourier_tv}) the formulation of the entropy current density, $J_s$, in the presence of thermal conductivity can be made:

\begin{equation}  \label{entropy_current}
J_s = \frac{J_q}{T} = -\frac{\lambda \nabla T}{T} = -\frac{\lambda}{T} \nabla T .
\end{equation}

At this point we take the form of the quantized thermal conductivity given by Eq. (\ref{kvantalt_termikus_vezetokepesseg}). Similarly to $\Lambda$

\begin{equation}
\Lambda_s = \frac{\Lambda}{T} = \frac{{\pi}^{2} k_{B}^{2}}{3h} = 9.46 \cdot 10^{-13} \frac{\text{J/K}}{\text{K s}} 
\end{equation}

can be introduced as the quantized entropy conductance. This denotes the entropy flow per unit time and unit temperature. For a given temperature difference of $dT$ the entropy current is

\begin{equation}
I_S = -\Lambda_s dT.
\end{equation}

Recalling the relation from Eq. (\ref{temp}) --- and neglecting the $D \nabla^2 T$ term --- yields

\begin{equation}
dT = T - T_0 \sim -\frac{\partial\varphi}{\partial t}.
\end{equation}

In the present form the expression is integrable, by which the transferred entropy $S_{tr}$ can be expressed by the potential $\varphi$ as

\begin{equation}  \label{transferred_entropy}
S_{tr} = \Lambda_s (\varphi - \varphi_0) = \frac{{\pi}^{2} k_{B}^{2}}{3h}(\varphi - \varphi_0).
\end{equation}

At this point it is noted that the potential difference drives the system to equilibrium, and it leads to entropy change. This gives us a deeper meaning that the coefficient $\Lambda_s$ is the quantum of entropy conductance. \\
 
On the other hand, if $dT$ is related to the transmission of an energy packet, then using the relation of $\varepsilon = k_{B} dT$ yields that the entropy current can be formulated as

\begin{equation}
I_S = \frac{\Lambda_s}{k_{B}} \varepsilon = \frac{{\pi}^{2} k_{B}}{3h} \varepsilon .
\end{equation}

If the energy package is a single quantum with the energy of $\varepsilon = h \nu$ (e.g. a phonon or a photon) then the entropy current carried by it is

\begin{equation}  \label{entropia_aram}
I_S = \frac{{\pi}^{2} k_{B}}{3} \nu ,
\end{equation}

where $\nu$ is the frequency. \\

To proceed, it is required to formulate the entropy production density upon thermal transfer \cite{groot}

\begin{equation}
\sigma = J_q {\nabla} \frac{1}{T} = \lambda \left( \frac{\nabla T}{T} \right)^2 .
\end{equation}

This can be readily reformulated in quantized form using Eqs. (\ref{entropy_current}) and (\ref{kvantalt_hovezeto_kepesseg}). The obtained result is

\begin{equation}
\sigma = \frac{J_s^2}{\lambda} = \frac{1}{T} \frac{\pi^2 \varepsilon^2}{3h} \frac{1}{A L}.
\end{equation}

Multiplying this equation by the volume $V = A L$ and introducing the entropy production $\Sigma = \sigma V$ of the considered system, the entropy production is thus

\begin{equation}
\Sigma = \frac{1}{T} \frac{\pi^2 \varepsilon^2}{3h}.
\end{equation}

If the energy transfer can be expressed by an energy packet (or a quasi particle) with frequency $\nu$, then for the entropy production the following relation holds:

\begin{equation}  \label{entropia_prod}
\Sigma = \frac{1}{T} \frac{\pi^2}{3} h \nu^2 .
\end{equation}

It is remarkable that the expression relates to the square of frequency $\nu$.

\section{Examples and applications}

The examples are detailed below to apply the elaborated framework of quantized thermodynamic condutances.

\subsection{Entropy increase during a single quantum transfer}

Let us consider two subdomains $\fbox{1}$ and $\fbox{2}$ with temperatures of $T_1$ and $T_2 < T_1$, respectively. Between the two subdomains a $h\nu$ energy transfer is taken into account. The energy packet is created in subdomain $\fbox{1}$, by which undergoes an entropy production of 

\begin{equation}
\Sigma_{1} = -\frac{1}{T_1} \frac{\pi^2}{3} h \nu^2.
\end{equation}

The negative sign is due to the formation of the quantum. The generated wave packet leaves the domain $\fbox{1}$, which has an entropy current of

\begin{equation}
I_{S_1} = -\frac{{\pi}^{2} k_{B}}{3} \nu
\end{equation}

originated from the domain $\fbox{1}$. Thus the total entropy decrease of the domain $\fbox{1}$ holds

\begin{equation}
\frac{dS_1}{dt} = -\frac{{\pi}^{2}k_{B}}{3} \nu -\frac{1}{T_1} \frac{\pi^2}{3} h \nu^2 .
\end{equation}

The energy packet arrives to the domain $\fbox{2}$, which yields an entropy current income of

\begin{equation}
I_{S_2} = \frac{{\pi}^{2} k_{B}}{3} \nu .
\end{equation}

On the other hand, the energy packet is spread in the volume causing an entropy production 

\begin{equation}
\Sigma_{2} = \frac{1}{T_2} \frac{\pi^2}{3} h \nu^2
\end{equation}

during this dissipation process. Consequently, the total entropy increase of domain $\fbox{2}$ is: 

\begin{equation}
\frac{dS_2}{dt} = \frac{{\pi}^{2}k_{B}}{3} \nu + \frac{1}{T_2} \frac{\pi^2}{3} h \nu^2 .
\end{equation}

The total entropy increase in the volume containing $\fbox{1+2}$ is thus:

\begin{equation}
\frac{dS}{dt} = \frac{dS_1}{dt} + \frac{dS_2}{dt} =  \left( -\frac{1}{T_1} + \frac{1}{T_2} \right) \frac{\pi^2}{3} h \nu^2 > 0,
\end{equation}

satisfying the $2^{\textrm{nd}}$ law of thermodynamics, as expected. During the formation of the quantum, the entropy decreases in the subdomain $\fbox{1}$ at temperature $T_1$. During the absortion the entropy increases in the subdomain $\fbox{2}$ at $T_2$. Since $T_1 > T_2$, the net entropy change is positive. If $T_1 = T_2$, e.g. in the thermal equilibrium, no further entropy is produced. The entropy current is independent of the temperature, so the transfer process has no further contribution the entropy increase. It is possible that a cooler subdomain may emit a quantum to the hotter one, but the resulting entropy increase must be positive, so reciprocally, the hotter subdomain must emit a higher energy (higher frequency) packet to the cooler one. This is the quantized formulation of second law of thermodynamics.

\subsection{Spin-lattice relaxation}

The spin–lattice relaxation is such a mechanism in which the parallel component of the nuclear magnetic moment relaxes from a higher energy instabil equilibrium state to the thermodynamic equilibrium. In the initial condition the magnetic moment is antiparallel to the constant magnetic field and the temperature equals to the temperaure of the surroundings, of the surrounding thermal bath. During the relaxation process the energy difference of the Zeeman levels has to be considered

\begin{equation}
\Delta E = \varepsilon = \gamma \hbar B_0,
\end{equation}

where $\gamma$ is the gyromagnetic ratio, and $B_0$ is the external magnetic field. To apply the results obtained above it is necessary to express the relevant frequency expression, like the Larmor frequency, which is 

\begin{equation}
\omega = \gamma B_0 \,\,\, \textrm{or} \,\,\, \nu = \frac{1}{2\pi} \gamma B_0. 
\end{equation}

The entropy current of a single spin relaxation, using the expression in Eq. (\ref{entropia_aram}), can be formulated in the expression 

\begin{equation}
I_{S} = \frac{{\pi}^{2} k_{B}}{3} \nu = \frac{{\pi} k_{B} \gamma B_0}{6}.
\end{equation}

The entropy production during the relaxation process can be obtained by the application of Eq. (\ref{entropia_prod}) as

\begin{equation}
\Sigma = \frac{1}{T} \frac{\pi^2}{3} h \nu^2 = \frac{1}{T} \frac{1}{12} h \gamma^2 B_0^2 .
\end{equation}

These results may be useful to understand, in general, the spin relaxation, basic thermodynamic relations in spintronics to achieve the minimal loss of spin-waves \cite{qin2021}, and in magnetic resonances \cite{csosz2020}, in the study of magnetic storage systems or quantum computing. Similar considerations can be made for any process with relaxation or interaction with light, e.g. photoluminescence.

\section{An additional consequence of the least action principle}

Let us turn back to the action principle of heat conduction. We could see that the action 

\begin{equation}  \label{action}
\tilde{S}(t) = \int\limits_{0}^{t} \frac{1}{2} (T-T_0)^{2} dt
\end{equation}

is the extremum (minimum) of the equalization process, i.e., it is minimal for the realizable motion. As it can be realized from Eq. (\ref{transferred_entropy}) pertains to the entropy transfer, thus the transferred entropy should be minimum for the realistic motion. Furthermore, this also means that the transferred energy during time $t$ is also related to the minimal entropy conductance. The factor, $\Lambda_s$, appeared in Eq. (\ref{transferred_entropy}) and multiplied by the action in Eq. (\ref{action}) returns a quantity which has the units of energy. Considering the aforementioned reasoning, the obtained quantity yields

\begin{equation}
\tilde{E} = \Lambda_s \tilde{S}(t) = \frac{{\pi}^{2} k_{B}^{2}}{3h} \int\limits_{0}^{t} \frac{1}{2} (T-T_0)^{2} dt, 
\end{equation}

which corresponds to the transferred enegy during the process. Finally, we conclude that the above formulated action principle is equivalent with the minimal entropy production principle of time dependent (non-stationary) processes. The formulated relations bring us closer to both the understanding of entropy current conductance and, eventually, to the meaning of the Lagrangian. \\

\section{Summary}

It is increasingly essential to understand the irreversibility of quantum mechanical and quantized transport processes on a microscopic scale. It is pointed out that both quantized entropy current and entropy production can be introduced and interpreted during the transfer of a single energy quantum. This completes the thermodynamic description of the process, including the validity of the second law of thermodynamics. Integrating into the theoretical framework of the of least action principle on thermal propagation, it became apparent that this principle may express as the principle of minimum entropy production on a quantum scale. We believe that our work is useful in the field of quantum computing understating how information loss can occur and thus how it can be tackled. \\


\section{Acknowledgments} 
We acknowledge the support of the NKFIH Grant nos. K119442 and 2017-1.2.1-NKP-2017-00001. The research has also been supported by the NKFIH Fund (TKP2020 IES, grant no. BME-IE-NAT) based on the charter of bolster issued by the NKFIH Office under the auspices of the Ministry for Innovation and Technology.


\baselineskip 12pt

\bibliographystyle{ieeetr}

\end{document}